\documentclass[iop]{emulateapj}
\usepackage{amsmath}
\usepackage{natbib}

\submitted{Accepted to ApJ on March 29, 2015}

\begin{document}

\title{Two Populations of Old Star Clusters in the Spiral Galaxy M101 Based on $HST$/ACS Observations}
\shorttitle{Two Populations of Old Star Clusters in M101}

\author{Lesley A. Simanton\altaffilmark{1}, Rupali Chandar\altaffilmark{1}, and Bradley C. Whitmore\altaffilmark{2}}
%\email{lesley.simanton@rockets.utoledo.edu}

\altaffiltext{1}{Department of Physics and Astronomy, University of Toledo, Toledo, OH 43606}
\altaffiltext{2}{Space Telescope Science Institute, Baltimore, MD 21218}

%\email{rupali.chandar@utoledo.edu}
%\email{whitmore@stsci.edu}

\begin{abstract}

We present a new photometric catalog of 326 candidate globular clusters (GCs) in the nearby spiral galaxy M101, selected from $B$, $V$, and $I$ $Hubble$ $Space$ $Telescope$ Advanced Camera for Surveys images.  The luminosity function (LF) of these clusters has an unusually large number of faint sources compared with GCLFs in many other spiral galaxies.  Accordingly, we separate and compare the properties of ``bright" ($M_V < -6.5$) versus ``faint" ($M_V > -6.5$; one magnitude fainter than the expected GC peak) clusters within our sample.  The LF of the bright clusters is well fit by a peaked distribution similar to those observed in the Milky Way (MW) and other galaxies.  These bright clusters also have similar size ($r_{\text{eff}}$) and spatial distributions as MW GCs.  The LF of the faint clusters, on the other hand, is well described by a power law, $dN(L_V)/dL_V \propto L_V^{\alpha}$ with $\alpha = -2.6 \pm 0.3$, similar to those observed for young and intermediate-age cluster systems in star forming galaxies.  We find that the faint clusters have larger typical $r_{\text{eff}}$ than the bright clusters, and have a flatter surface density profile, being more evenly distributed, as we would expect for clusters associated with the disk.  We use the shape of the LF and predictions for mass-loss driven by two-body relaxation to constrain the ages of the faint clusters.  Our results are consistent with two populations of old star clusters in M101: a bright population of halo clusters and a fainter, possibly younger, population of old disk clusters. 


\vspace{10 mm}

\end{abstract}

\section{Introduction}

Ancient ($\ga10$~Gyr) star clusters formed during the early assembly of most galaxies, and therefore give insight into the broad formation history of their hosts.  Elliptical and lenticular galaxies host red, metal-rich globular clusters (GCs) believed to be associated with their bulges, and blue, metal-poor GCs believed to be associated with their halos.  It is not yet clear whether these different populations have different ages in all early-type galaxies, but an age difference has been observed in several \citep{par12}.  Spiral galaxies also form metal-poor halo and metal-rich bulge/thick disk clusters, although the fraction of red-to-blue GCs is typically lower than found in similar mass early-type galaxies.  \citet{min95} and \citet{cot99} showed that metal-rich Milky Way (MW) GCs in the inner regions of the Galaxy are associated with the bulge.  

In addition to ancient GCs, spiral galaxies form younger clusters with a large range of ages in their disks.  In the MW, young ($<100$~Myr) open clusters are confined to the thin disk and old ($>3$~Gyr) open clusters are found in the thick disk \citep{por10}.  In fact, \citet{kha13} find a smoothly increasing dispersion in the distance of MW star clusters from the Galactic plane with increasing cluster age (see their Figure~5).

The GC luminosity functions (LFs) of both elliptical and spiral galaxies, including the MW, are observed to have a peaked shape with a ``turnover" caused by the earlier disruption of lower mass clusters due to ``evaporation" of stars by two-body relaxation in a tidal field \citep{fal01}.  Small variations in the peak luminosity and width of GCLFs are believed to come from possible differences in cluster ages, cluster densities, or the strength of the tidal field \citep{vil10}.  Despite the known small variations, the turnover of GCLFs is nearly universal, which has led to its use as a standard candle for distance determinations \citep[p. 223]{har01}.  

It is therefore important to study cases where the GCLF does not follow the ``universal" pattern.   An interesting example is the so-called ``faint fuzzies" that have been found in the lenticular galaxies NGC 1023 and NGC 3384 \citep{bro02,lar00}, lenticular and elliptical galaxies in the Virgo cluster \citep{pen06_2}, and M51 and its companion \citep{sch07,hwa06}.  These clusters are quite extended with effective radii ($r_{\text{eff}}$) of 7-15~pc and ages upwards of 7-8~Gyr.  Despite their intermediate/old ages, their LF is observed to continue to rise beyond the expected turnover luminosity.  Intermediate age clusters found in a few early-type galaxies may be responsible for discrepancies between the distances measured from GCLFs and those measured from surface brightness fluctuations \citep[p. 281]{ric03}.

Faint fuzzies also have $r_{\text{eff}}$ more than twice that of typical GCs in most ellipicals and spirals.  \citet{bro02} proposed that perhaps faint fuzzies only form in the environment found in lenticuler galaxies, or that perhaps they were accreted along with a host dwarf galaxy.  However, \citet{chi13} used kinematics and spatial comparisons with planetary nebulae and \ion{H}{1} to conclude that faint fuzzies may simply be akin to old open disk clusters.  They suggest that the reason faint fuzzies are observed in lenticulars but not in spirals is that these faint clusters are quite difficult to pick out against the strong structure and variable luminosity caused by star formation in spiral disks.  

Thus far, most detailed studies of GCs have focused on early type galaxies (e.g. \citet{pen06,vil10}), the MW, M31, and a handful of other, mostly bulge dominated systems (e.g. \citet{gou03}).  Late-type spirals, such as M101, have not been as thoroughly studied.  Previously, \citet{cha04} studied the GC systems in five spiral galaxies using $Hubble$ $Space$ $Telescope$ ($HST$) Wide-Field Planetary Camera 2 (WFPC2) observations and compared with the known distribution of GCs in the MW.  They found for M101 and NGC 6946 that the LF continues to rise beyond the expected GCLF turnover; however, this conclusion was based on a small number of clusters resulting from the partial coverage of each galaxy.  \citet{bar06} studied a larger sample of M101 clusters (1715 clusters with $m_V < 23$), but they focused on analyzing ``blue" clusters (($B-V$)$_0<0.45$) rather than the older, redder clusters we seek to study here.  In this paper, we use the same $BVI$ $HST$ Advanced Camera for Surveys (ACS) observations as \citet{bar06} to examine the LF, colors, sizes, and spatial distributions of the red clusters. 

The rest of the paper is organized as follows:  we describe the observations, selection criteria and methods, and completeness in $\S2$.  In $\S3$, we show the luminosity distribution, color-color plot, color histograms, color magnitude diagram (CMD), sizes, and spatial distribution of our cluster candidates, and we discuss the results in $\S4$.  Finally in $\S5$, we list our conclusions and describe ongoing work on the M101 cluster system.

\section{Observations, Cluster Selection, and Completeness}

\subsection{Observations}

\begin{figure}[htp]
\begin{center}
\includegraphics[width=0.48\textwidth]{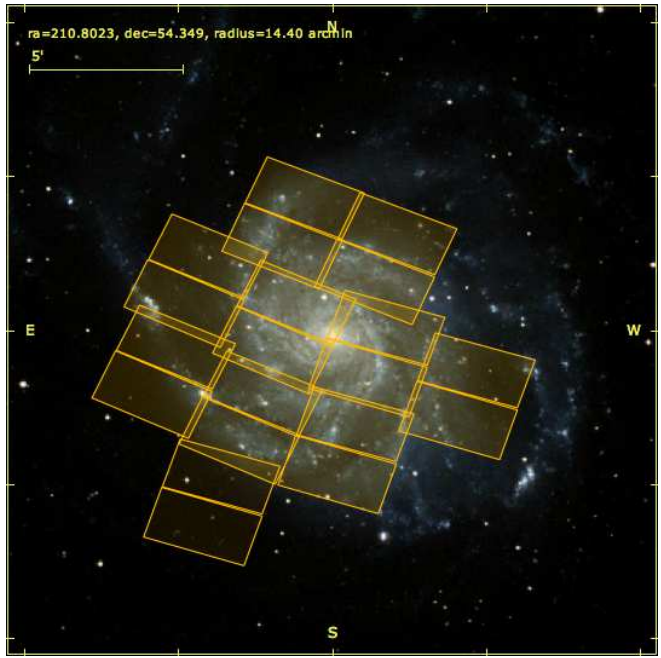}
\caption{Ground-based optical image of M101 showing the location of all 10 $HST$ ACS/WFC fields used in this work.  The $5'$ scale bar is equivalent to $\sim$9.3 kpc.}
\label{fields}
\end{center}
\end{figure}

Ten pointings within M101 were taken with the $HST$/ACS Wide Field Channel (WFC) in November of 2002 (Program ID:  9490, PI:  K. Kuntz) using the $F435W$ ($B$), $F555W$ ($V$), and $F814W$ ($I$) filters\footnote{Based on observations made with the NASA/ESA Hubble Space Telescope, and obtained from the Hubble Legacy Archive, which is a collaboration between the Space Telescope Science Institute (STScI/NASA), the Space Telescope European Coordinating Facility (ST-ECF/ESA) and the Canadian Astronomy Data Centre (CADC/NRC/CSA).}.   Each field covers $3.3' \times 3.4'$ (see Fig.~\ref{fields}).   Assuming a distance to M101 of $6.4 \pm 0.2$ (random) $\pm 0.5$ (systematic)~Mpc \citep{sha11}, each field covers a 6.1 $\times$ 6.3~kpc$^2$ region.  Images were processed through the HLA MultiDrizzle Pipeline Version 1.0 \citep{koe02}, which includes bias subtraction, cosmic ray rejection by combining two sub-exposures from CR-Split observing, dark subtraction, flat fielding, and drizzling.

We detect sources, which include star clusters, bright individual stars, and background galaxies, in the $V$ band image using the IRAF task DAOFIND.  We detect $\sim$383,000 total sources and measure their brightnesses within circular apertures varying from 0.5 to 5 pixels with the background estimated within annuli of 7 to 13 pixels using the PHOT task within IRAF.  We determine empirical aperture corrections out to 10 pixels from the curves of growth measured for 25 isolated clusters, and also apply an additional $-0.107$, $-0.092$, and $-0.087$~mag correction from 10 pixels to infinity for $B$, $V$, and $I$ respectively \citep{sir05}.  These aperture corrections are added to the measured photometry to obtain instrumental magnitudes.  The apparent magnitude ($m_V$) for each source in the VEGAMAG system is found by applying the following zero points:  $F435W = 25.791$, $F555W = 25.738$, and $F814W = 25.533$ \citep{boh07,mac07}.\footnote{\url{http://www.stsci.edu/hst/acs/analysis/zeropoints}}

\subsection{Cluster Selection}

In order to separate ancient star cluster candidates from our full catalog, we made a series of automated selective cuts followed by visual inspection of each object.  We use the following automated selection criteria:

\begin{enumerate}
\item Clusters brighter than $m_V < 24.75$ where $m_V$ was measured within 3 pixel apertures (i.e. no aperture correction yet applied) to ensure high signal-to-noise ratio.
\item Concentration index (CI) $>  1.15$ where CI is the difference between $m_V$ measured within 1 pixel and 3 pixel apertures to eliminate point sources; point sources have CI values that peak around $1.00$ with a standard deviation of $0.06$.
\item Colors, i.e. $0.55 < B-V < 2.0$ and $0.75 < V-I < 2.0$, similar to those of Galactic GCs. 
\end{enumerate}

Because CI is a crude measure of object size, we also use the BAOlab/ISHAPE software \citep{lar99} to measure the FWHM of each object.  ISHAPE fits profiles to each candidate source to determine its FWHM (along with other parameters such as ellipticity).  The light profile is a convolution of the point spread function (PSF) with a user determined function representing the spread in light from a cluster's non-point-like size.  We determined the PSF of each of our fields by visually selecting $\sim$40-50 isolated stars in each field.  We choose a King profile \citep{kin62} with a concentration parameter (ratio of the tidal radius to the core-radius) of 30 to represent the cluster-like light profile convolved with the PSF for the ISHAPE fitting.  

All objects with a measured FWHM $< 0.2$~pixels ($r_{eff}\approx0.46$~pc) from ISHAPE were removed from the catalog to further eliminate stars.  We then plotted FWHM versus CI and determined a three piece-wise linear fit to the relationship between CI and FWHM.  Upon visual inspection, objects falling outside of a perpendicular distance of $\sim0.1$ from the line only consisted of contaminants (usually faint, distorted ``patches" that cannot be classified, noise, crowded point sources, etc).  Therefore, only objects within a perpendicular distance of 0.1 from the CI-FWHM fits were kept. 

\begin{figure}[htp]
\begin{center}
\includegraphics[width=0.48\textwidth]{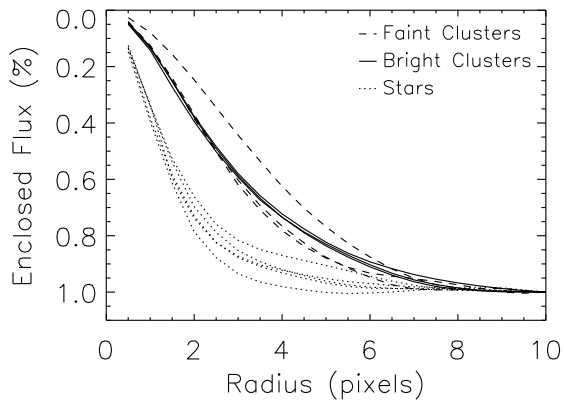}
\caption{Radial profile showing the enclosed percentage of flux within a given radius (in pixels) for three faint clusters (dashed lines), three bright clusters (solid lines), and five stars (dotted lines).  The point sources have significantly steeper profiles.}
\label{radprof}
\end{center}
\end{figure}

Finally, obvious background galaxies, chance superpositions, and other contaminants were eliminated via visual inspection.  Figure~\ref{radprof} shows a few typical radial profiles examined during the visual inspection of six selected candidate clusters and five stars (not included in the catalog).  Note the clear difference between stars and the relatively well-resolved clusters. 

\begin{figure}[htp]
\includegraphics[width=0.48\textwidth]{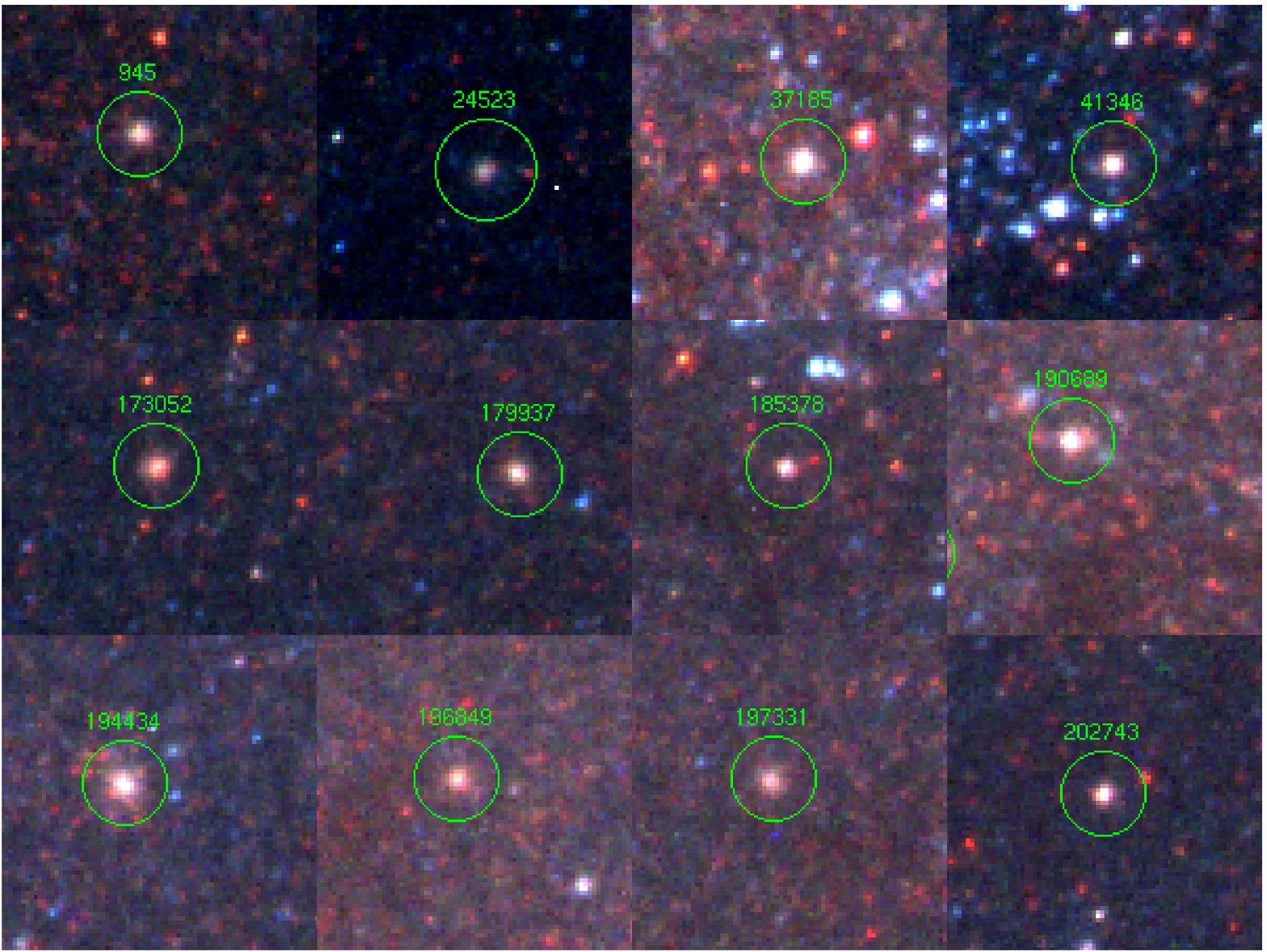}
\includegraphics[width=0.48\textwidth]{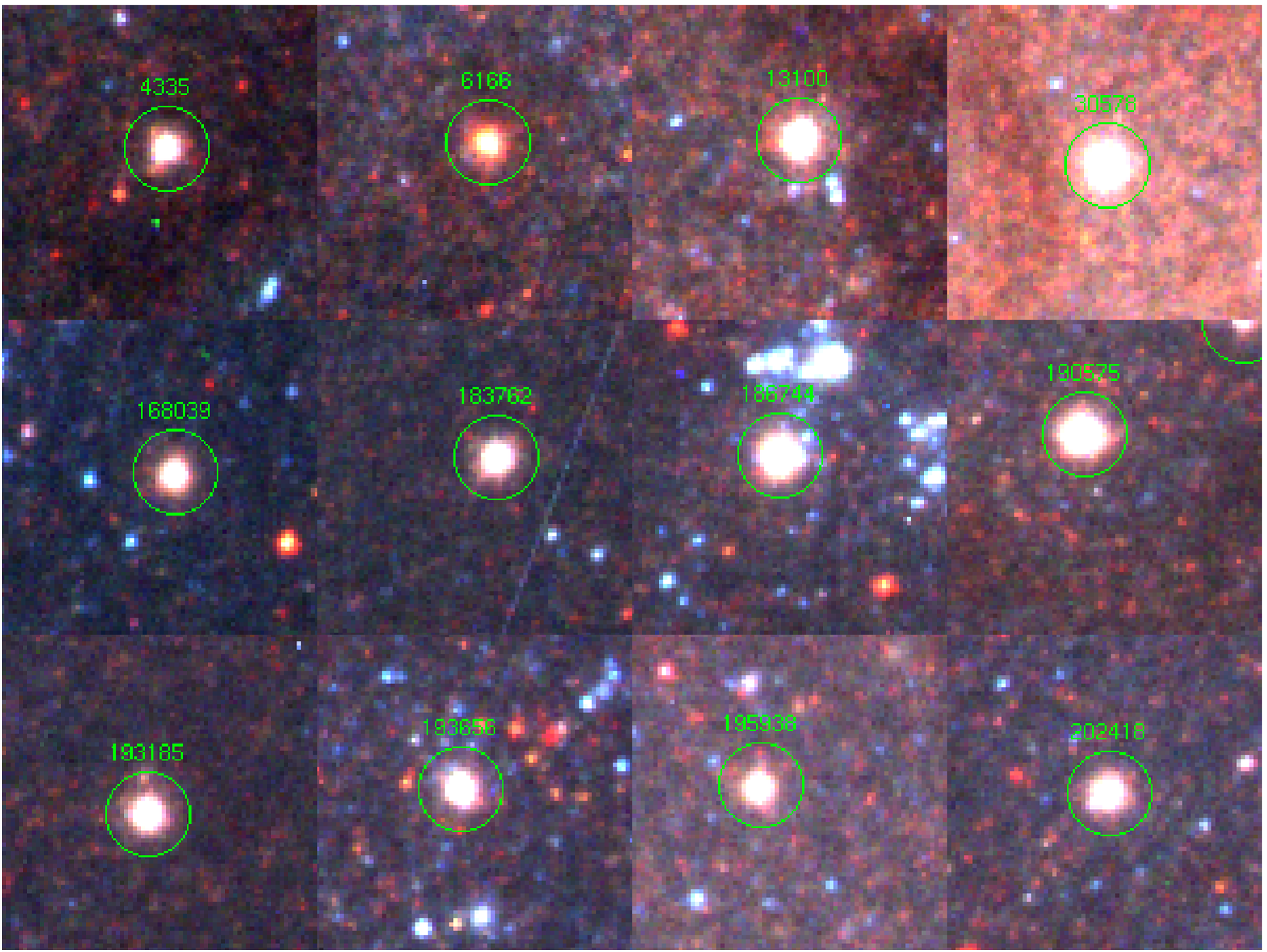}
\caption{$BVI$ color images of typical clusters from faint (top panel) and bright (bottom panel) cluster groups.  Each panel is $\sim3.7''$, or $\sim$110 pc on a side.}
\label{stamps}
\end{figure}

The final catalog consists of 326 candidate clusters (see Fig.~\ref{stamps} and Table~\ref{catalog}) with magnitude, color, and size measurements (six of these do not have FWHM measurements due to ISHAPE fit errors).  For simplicity, we refer to the cluster candidates as ``clusters" throughout the rest of the paper.

%%add full catalog
\begin{deluxetable*}{lccccccc}
\tablecolumns{8}
\tablewidth{0pc}
\tablecaption{M101 Old Star Cluster Catalog\tablenotemark{a}}
\tablehead{\colhead{ID} & \colhead{$\alpha_{2000}$~(hms)} & \colhead{$\delta_{2000}$~(\arcdeg\arcmin\arcsec)} & \colhead{$V$} & \colhead{$B-V$} & \colhead{$V-I$} & \colhead{$B-I$} & \colhead{$r_{\text{eff}}$ (pc)}}
\startdata
945 & 14 3 16.68 & 54 19 38.04 & -5.89 & 0.68 & 1.08 & 1.77 & 5.3 \\
3072 & 14 3 22.28 & 54 19 50.40 & -5.76 & 0.64 & 0.97 & 1.61 & 3.8 \\
4335 & 14 3 17.58 & 54 19 55.68 & -7.35 & 0.72 & 0.98 & 1.70 & 3.6 \\
6166 & 14 3 18.09 & 54 20 3.09 & -7.05 & 1.28 & 1.75 & 3.02 & 2.4 \\
9429 & 14 3 15.44 & 54 20 16.08 & -6.77 & 0.93 & 1.34 & 2.27 & 2.0 \\
9936 & 14 3 16.31 & 54 20 18.48 & -5.96 & 0.79 & 1.16 & 1.95 & 7.3 \\
10101 & 14 3 18.62 & 54 20 19.33 & -6.30 & 0.68 & 1.04 & 1.72 & 7.6 \\
10379 & 14 3 23.53 & 54 20 20.59 & -6.49 & 0.76 & 1.21 & 1.97 & 4.7 \\
10633 & 14 3 25.18 & 54 20 21.79 & -8.83 & 0.80 & 1.19 & 1.99 & 2.3 \\
10654 & 14 3 20.35 & 54 20 21.89 & -5.89 & 0.84 & 1.19 & 2.03 & 7.7 \\
\enddata
\tablenotetext{a}{This table is available in its entirety in machine-readable form.}
\label{catalog}
\end{deluxetable*}

\subsection{Completeness}

To evaluate the completeness of our catalog of clusters, we generate artificial clusters, add them to the M101 images, and re-run the detection and cluster selection methods described in $\S2.1$ and $\S2.2$.  We generate 4000 artificial clusters using the BAOLAB task MKCMPPSF, which convolves a PSF with a user defined function, in this case KING30 profiles with two different input FWHM: 1.0 and 2.0 pixels (the motivation for these sizes is given in $\S3.3$).  We then use MKSYNTH to randomly place the artificial clusters in one of the images, where the magnitude range matches that of the real clusters.  We detect sources with DAOFIND using the same parameters as for the real clusters, measure photometry with PHOT, size measurements with ISHAPE, and run the automated cluster selection criteria and CI-FWHM relation cut (see $\S2.2$).  

\begin{figure}[htp]
\includegraphics[width=0.48\textwidth]{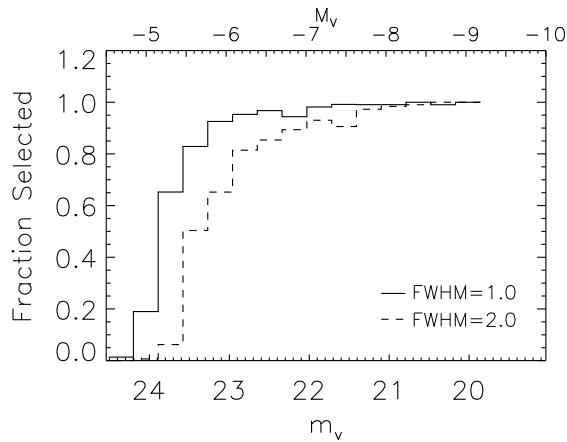}
\caption{Fraction of selected artificial clusters versus $m_{v}$ for FWHM$=1.0$ (solid line) and 2.0 (dashed line) pixels.}
\label{maghisto_art}
\end{figure}

Figure~\ref{maghisto_art} shows the completeness fraction as a function of apparent magnitude.  Both sizes of artificial clusters show a decline in completeness as brightness decreases with the more diffuse clusters declining at slightly brighter magnitudes.  Approximately $80\%$ of the artificial clusters brighter than $m_V = 23.0$ make it through our selection pipeline, and approximately $50\%$ of those brighter than $m_V\approx23.6$. 

\begin{figure}[htp]
\includegraphics[width=0.48\textwidth]{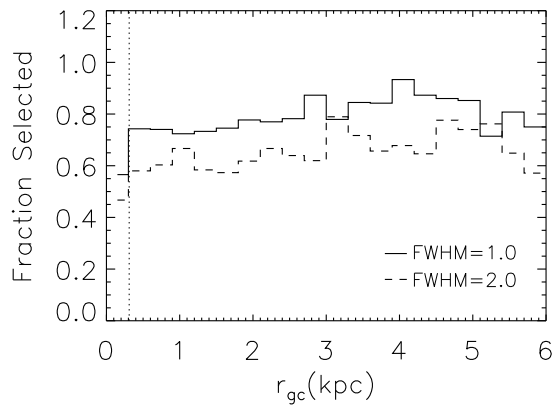}
\caption{Fraction of selected artificial clusters versus $r_{\text{gc}}$ for FWHM$=1.0$ (solid line) and 2.0 (dashed line) pixels.  The dotted line represents the innermost region excluded from the fits in Figure \ref{surfacedensity} (see $\S3.4$), which is the only significant drop in completeness as a function of distance.}
\label{disthisto_art}
\end{figure}

Although the clusters are randomly placed in the M101 images, we ensured that 2000 are located within galactocentric distance $r_{\text{gc}}\approx2$~kpc and 500 within $r_{\text{gc}}\approx500$~pc to ensure that the completeness of the inner regions of M101 are thoroughly tested.  Figure~\ref{disthisto_art} shows the completeness fraction as a function of $r_{\text{gc}}$.  While the completeness is slightly better for more compact clusters, there is no significant change in the completeness fraction with $r_{\text{gc}}$ except within the innermost $\sim 300$~pc (dotted line in Fig.~\ref{disthisto_art}).

\section{Results and Analysis}

\subsection{Cluster Luminosity Distribution}

\begin{figure}[htp]
\includegraphics[width=0.48\textwidth]{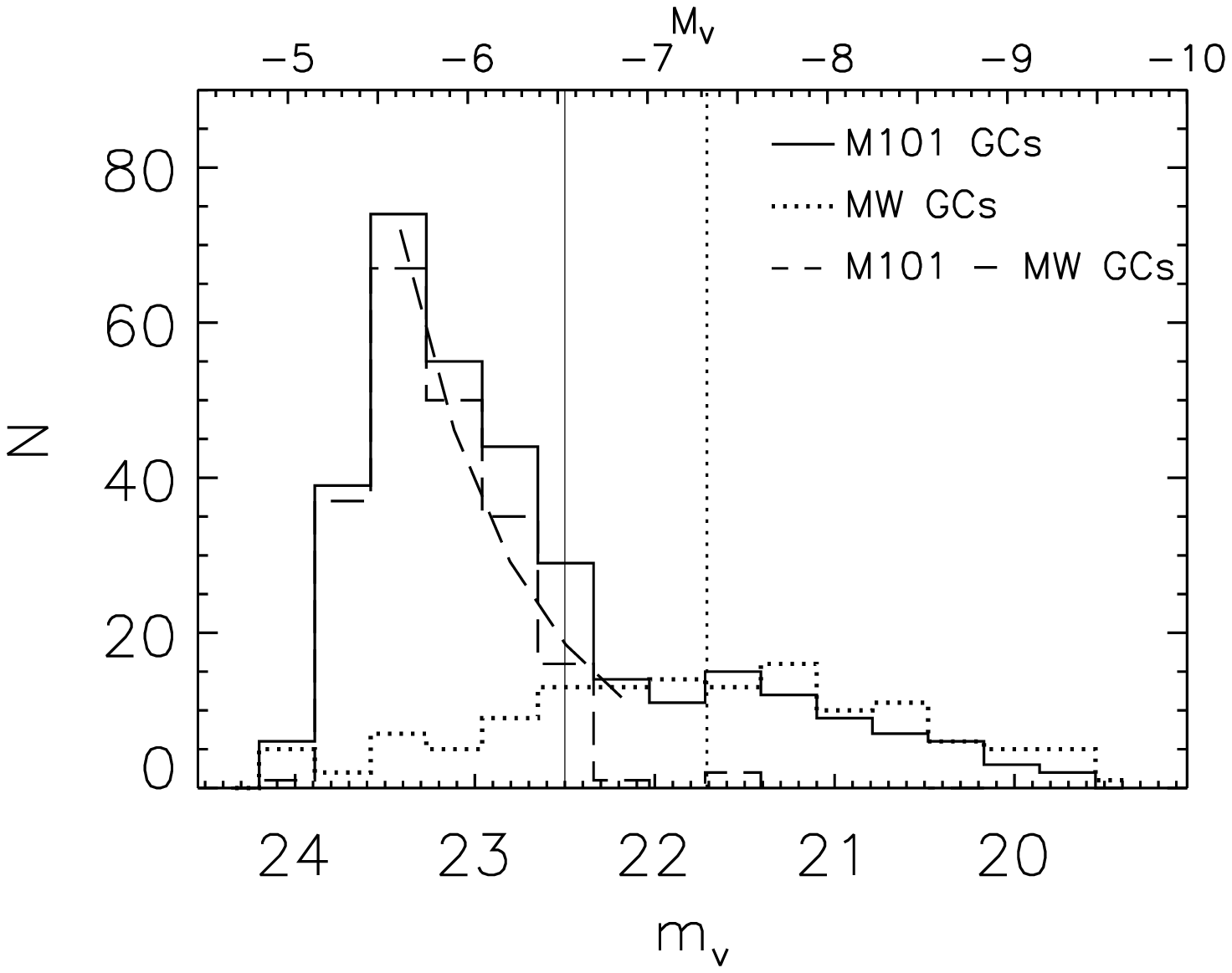}
\includegraphics[width=0.48\textwidth]{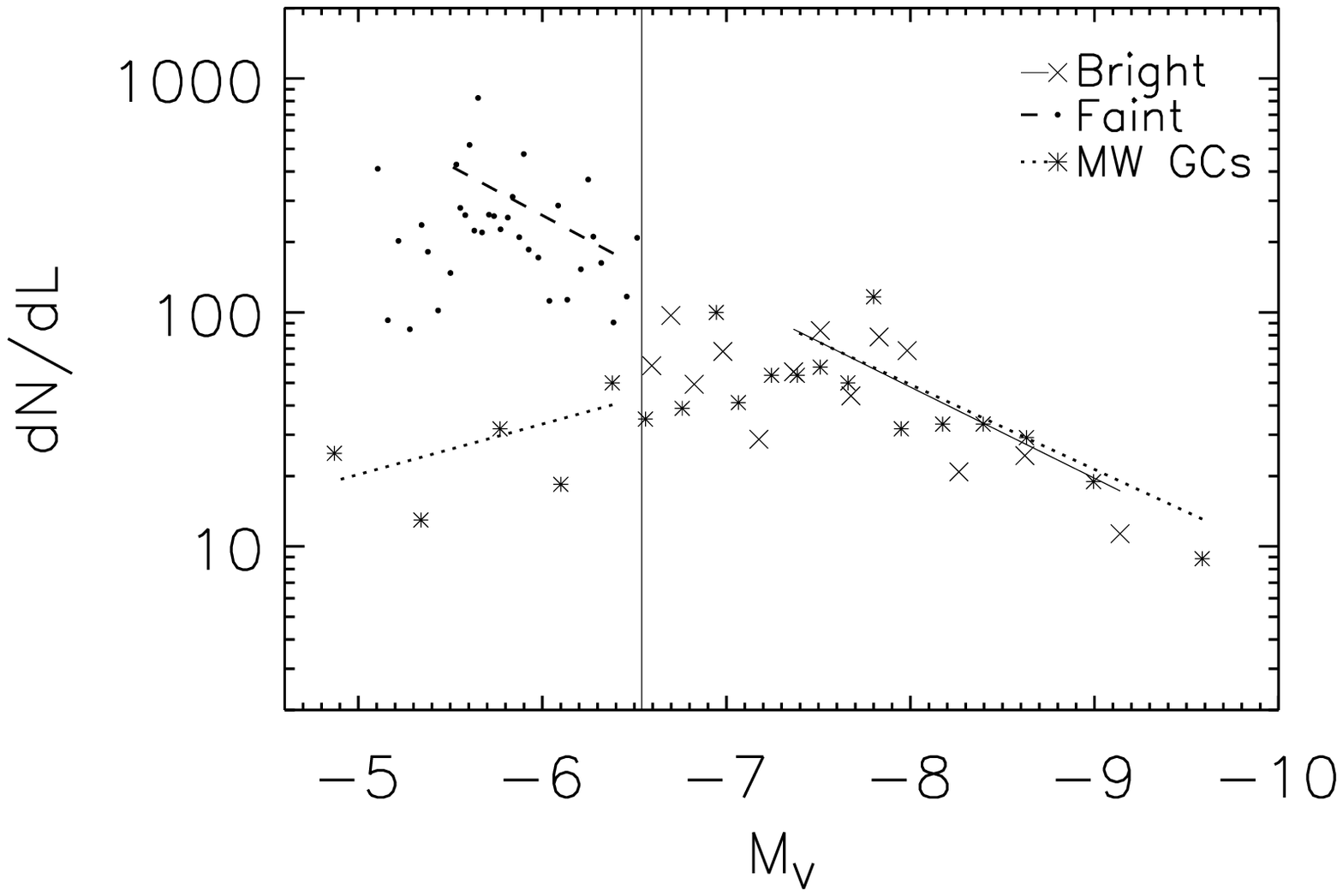}
\caption{{\bf Top:} magnitude distribution (top) for our M101 cluster sample (solid line), MW GCs (dotted line), and the difference between the M101 clusters and the MW GCs (dashed line).  A power law fit (dashed line) to the dashed line histogram for the bins brighter than the completeness limit gives $\alpha = -2.60 \pm 0.26$.  The solid vertical line represents $m_V = 22.5$ ($M_V = -6.54$), which is where we choose to divide our sample into ``faint" and ``bright" clusters (see text).  The dotted vertical line represents the peak of the MW GC distribution. {\bf Bottom:} luminosity distribution for our bright clusters (crosses), faint clusters (solid circles), and MW GCs (asterisks) with constant number binning.  Fits to subranges of each group are shown with the same line styles as the top panel.}
\label{lumhisto}
\end{figure}

Figure~\ref{lumhisto} shows the magnitude and luminosity distributions for the M101 cluster sample alongside those for MW GCs (taken from the \citet{har96} catalog).  The top panel shows a histogram with equal size magnitude bins, while the lower panel shows the luminosity distribution with an equal number of clusters (seven) in each bin.  In the magnitude distribution, the M101 clusters have a similar shape to those in the MW (see $\S4$ for further discussion) at the bright end, but a drastically different one at the faint end.  

In the rest of this section, we compare the results of fitting different portions of the cluster luminosity distributions.  These fits are performed with a simple least squares regression in $log N$-$m_V$ space.  The slope, $a$, from these fits can be converted into the true power law index $\alpha$ for the LF $dN(L_V)/dL_V \propto L_V^{\alpha}$ by $\alpha=-(2.5a + 1)$.  We first compare the results of fitting the bright end of the M101 and MW GC distributions (see the bottom panel of Fig.~\ref{lumhisto}).  The fit range is $M_V \le -7.5$, approximately $0.2$~mag brighter than the peak in the MW GCLF \citep{ash95}.  The best fits shown in Figure~\ref{lumhisto} are $\alpha_{M101,bright}=-1.97 \pm0.14$ and $\alpha_{MW,bright}=-1.91 \pm0.11$.  The standard deviation for 16 different binnings and ranges are $\sim0.16$ for both the M101 and MW data, comparable to the uncertainties. 

For the faint clusters, we use a magnitude range of $-5.4 \gtrsim M_V \gtrsim -6.5$.  Here, completeness is a factor, and fainter than $M_V \approx -5.4$, the LF declines steeply, indicating the probable completeness limit, which matches well with the magnitude at which the completeness fraction of the artificial clusters drops below $50\%$ (see Fig.~\ref{maghisto_art}).  We fit the MW GCs in the full faint range $M_V > -6.54$ ($m_V > 22.5$) since the completeness limit is not a concern for the \citet{har96} catalog.  The slopes for these fits are $\alpha_{M101,faint}=-2.03 \pm 0.05$ and $\alpha_{MW,faint}=-0.46 \pm 0.17$\footnote{Note that the value for $\alpha$ for the faint MW GCs is negative, while $a$, which is shown in the lower panel of Fig.~\ref{lumhisto}, is positive.}.  The standard deviation for 16 different binnings and ranges (including more conservative completeness limits brighter than $M_V\approx-5.4$) are $\sim0.25$ and $\sim0.22$ respectively.  The shapes of the faint ends of the M101 and MW GC distributions are very different.  Note that any faint clusters not identified in our M101 sample due to incompleteness will only steepen $\alpha_{M101,faint}$ and increase the discrepancy between the M101 and MW GC distributions.

Statistical tests confirm the fit results for the luminosity distribution of M101, that the bright end is similar to that in the MW, but the faint end is quite different.  We compare the shapes of the M101 and MW luminosity distributions using the two-sided Kolmogorov-Smirnov (KS) test and the two-sided Cramer-von Mises (CvM) test.  Both tests accept the null hypothesis ($p$-values $> 0.1$) that the M101 luminosity distribution and the MW luminosity distribution are drawn from the same distribution at the bright end $M_V < -6.54$, and very strongly reject the null hypothesis ($p$-values $< 0.01$) at the faint end $M_V > -6.54$.  

In fact, the fits do not produce $p$-values $<0.05$ (strong rejection) until the samples include clusters fainter than $M_V > -6.54$.  Therefore, we divide our sample into two groups throughout the rest of the paper, where ``bright" clusters have $M_V<-6.54$ ($m_V<22.5$) and ``faint" clusters have $M_V>-6.54$ ($m_V>22.5$).  There are 90 clusters in the bright group and 236 clusters in the faint group. 

We perform a final fit to the faint clusters after subtracting the MW GC histogram from the M101 cluster histogram (see the dashed line in the top panel of Figure~\ref{lumhisto}).  No normalization of the MW GC histogram is performed prior to subtraction because the bright ends of the distributions are well matched (see Fig.~\ref{lumhisto}).  We find $\alpha=-2.60 \pm 0.26$ (uncertainties from Poisson errors, i.e. $1/\sqrt{N}$) in the range $m_V=23.58$ to $m_V=22.03$.  Nine different combinations of bin sizes and data ranges (always brighter than the completeness limit, $m_V\approx23.65$) give a median $\alpha =-2.57$ and a standard deviation of $\sim0.18$.

\subsection{Cluster Colors and Luminosities}

In this section, we compare the colors and luminosities of old star clusters in M101 with those of cluster populations in other galaxies.  The ages and metallicities of clusters older than $\sim1$~Gyr become degenerate, making it difficult to establish differences in age based only on integrated broad-band colors; the colors are a stronger indicator of metallicity than of age in this regime.  

\begin{figure}[htp]
\includegraphics[width=0.48\textwidth]{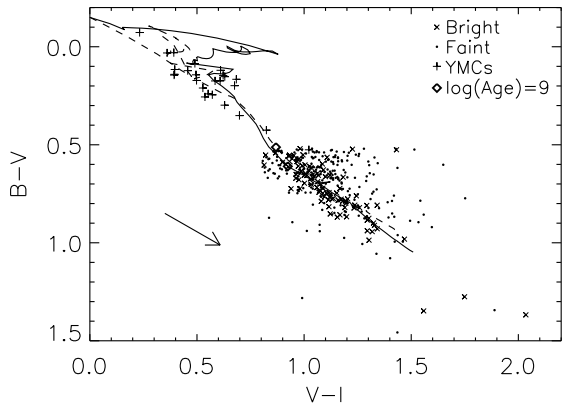}
\caption{Color index plot showing $B-V$ vs. $V-I$ colors for our full M101 cluster sample faint (solid circles) and bright (crosses) groups and the M101 YMCs with spectra (pluses).  The lines show the evolution of \citet{bru03} (updated 2006 data) SSP models with solar (solid line) and $Z=0.008$ (dashed line) metallicities.  The 1 Gyr age on each SSP track is indicated by an open diamond.  The M101 clusters follow both SSP tracks well with some spread most likely due to photometric uncertainties.  The arrow represents typical Galactic reddening for $A_{V}=0.5$.  Galactic reddening in the direction of M101 is very low ($E(B-V)<0.01$, \citet{cha04}), and we do not correct for it.}
\label{colors}
\end{figure}

For bright clusters, the median $B-V$, $V-I$, and $B-I$ colors are 0.72, 1.10, and 1.81, respectively, and for the faint clusters they are 0.65, 1.10, and 1.76.  Figure~\ref{colors} shows $B-V$ versus $V-I$ colors with two metallicity simple stellar population (SSP) tracks (solar and $Z=0.008$) from \citet{bru03} updated 2006 data (from private communication).  The cluster colors are approximately centered on the SSP tracks with some spread in the colors.  This spread is almost certainly due to photometric errors, based on a comparison between input and measured magnitudes for artificial clusters.  Furthermore, the spread in cluster colors around the SSP tracks increases for fainter clusters, as expected when photometric errors are the dominant source of uncertainty.  Therefore, the small differences between the median values for the faint versus bright cluster colors in $B-V$ and $B-I$ are unlikely to be strongly significant. 

\begin{figure}[htp]
\includegraphics[width=0.48\textwidth]{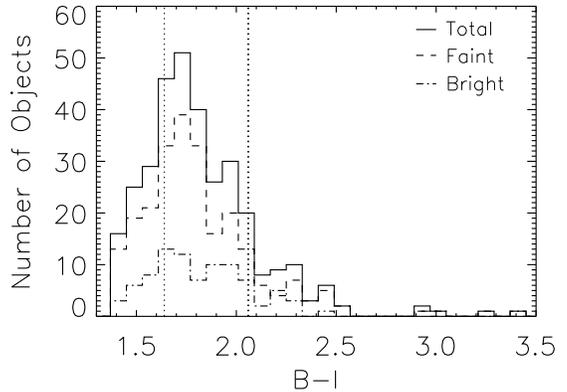}
\includegraphics[width=0.48\textwidth]{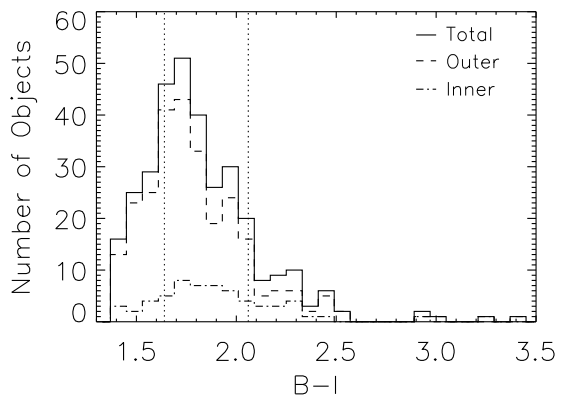}
\caption{{\bf Top:} $B-I$ color histogram with M101 clusters divided into bright (dashed-dotted line) and faint (dashed line) groups.  Vertical dotted lines are the mean blue and red peak $B-I$ colors for eight BCGs \citep{har06}.  {\bf Bottom:} same as above, except the clusters are now divided by distance with the dashed line representing clusters outside of 3 kpc from the center of the galaxy and the dashed-dotted line representing inside of 3 kpc.}
\label{colorhisto}
\end{figure}

\begin{figure}[htp]
\includegraphics[width=0.48\textwidth]{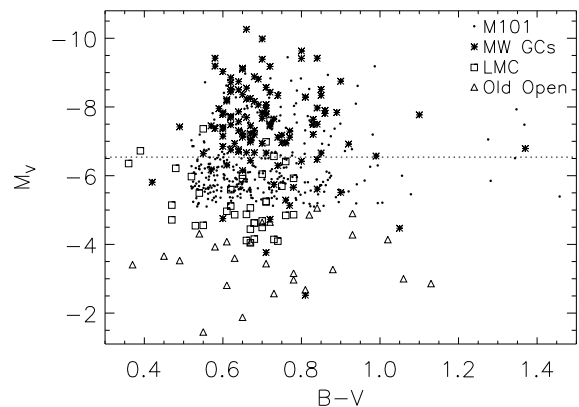}
\caption{$B-V$ color magnitude diagram showing our M101 sample (solid circles), MW GCs (asterisks), LMC intermediate age clusters (1-3 Gyr, open squares), and old open clusters (1-8 Gyr, open triangles).  The dotted line corresponds to the luminosity division at $M_V = -6.54$ imposed on our M101 sample (see $\S 3.1$).}
\label{CMD}
\end{figure}

The $B-I$ color has the largest wavelength baseline and is therefore best suited to revealing multiple peaks in the colors, if they exist.  Figure~\ref{colorhisto} shows $B-I$ histograms.  The top plot shows the bright and faint M101 cluster groups, and the bottom plot divides the clusters by distance, either inside or outside of 3~kpc from the center of M101 \citep{eva10}.  The faint clusters drive this peaked distribution; the bright clusters appear to have a more even-color distribution with no obvious peak.  Overlaid (dotted lines) are the mean blue and red peak $B-I$ colors of eight brightest cluster galaxies (BCGs) determined by \citet{har06}.  Note that we have not converted from the VEGAMAG system to the Johnson-Cousins system.  We estimate from Figure~21 in \citet{sir05} that with a conversion to the Johnson-Cousins system, the M101 $B-I$ colors would shift blueward by at most $\sim-0.03$.  It is then clear that the total cluster histogram strongly peaks close to the blue BCG peak, with a weak tail of clusters extending to the red BCG peak.  This red tail may be caused by reddening within the host galaxy itself for which we have not accounted.  In the bottom panel of Figure~\ref{colorhisto}, the clusters within 3~kpc of the center of M101 are slightly redder than those outside of 3~kpc, although they do not have a peak centered on the red peak of the BCGs.  The redder color could result from higher extinction or possibly higher metallicity of these more centrally located clusters, but we cannot distinguish between these two possibilities with the currently available data.

Figure~\ref{CMD} shows the CMD for our cluster sample with MW GCs, Large Magellanic Cloud (LMC) intermediate age clusters (1-3 Gyr, magnitudes and colors from \citet{bic96} and ages from \citet{muc07}; \citet{pia09}; \citet{gou11}; \citet{pal13}, and \citet{ker07}), and old open MW clusters (1-8 Gyr, magnitudes, colors, and ages from \citet{lat02}) included for comparison.  Again, note that a conversion to the Johnson-Cousins filters has not been applied to the M101 clusters; we estimate $B-V$ colors would shift redward by at most $\sim+0.04$ \citep{sir05}.  The M101 cluster sample, as a whole, has colors similar to the MW GCs, but with significantly more clusters fainter than $M_V=-6.54$.  It is of note that the MW GC sample plotted here is missing $B-V$ for 40 clusters, 34 of which are in the faint cluster region.  Even taking this into account, however, there are still many more M101 faint clusters (236) than MW GCs (53) in this region.   

The LMC intermediate age clusters also lie largely in the region of the CMD dimmer than $M_V=-6.54$, but within the $B-V$ color range of the M101 clusters.  Old open clusters in the MW occupy a similar color space as the LMC clusters, but with even fainter magnitudes.  The median colors for the faint M101 clusters, intermediate age LMC clusters, the old open clusters, and MW GCs fainter than $M_V=-6.54$ are $B-V = 0.65, 0.67, 0.71$, and $0.74$ respectively, with the M101 clusters shifting up to $+0.04$ with conversion to the Johnson-Cousins system.  All of these groups may have consistent $B-V$ colors as the MW GC median value is affected by the 34 missing values.

Also shown in Figure~\ref{colors} are the colors for 25 young massive clusters (YMCs, $\sim$100s Myr) that are discussed further in $\S4.1$; they are part of a spectroscopic study on clusters in M101 (to be released in a follow-up paper).  Their categorization as YMCs is derived from the strength of the Balmer lines seen in $Gemini$/GMOS spectra.  The YMC colors shown in Figure~\ref{colors}, however, are measured from the $HST$ $BVI$ images studied here with the same treatment as the rest of the cluster catalog.  It is clear that the YMCs overall are much bluer than the faint and bright populations with a median $B-V = 0.17$ and $V-I = 0.52$.

\subsection{Sizes}

\begin{figure}[htp]
\includegraphics[width=0.48\textwidth]{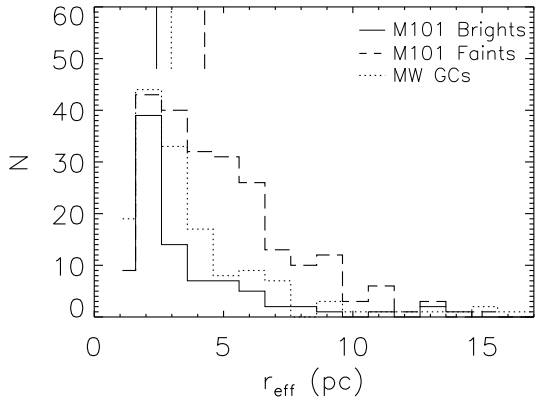}
\caption{$r_{\text{eff}}$ or half-light radius histogram for our M101 sample divided into bright (solid line) and faint (dashed line) clusters with the MW GCs (dotted line) also plotted.  The median values of the distributions are shown by the vertical lines at the top of the plot with their respective line styles for each group.  The median $r_{\text{eff}}$ of the bright M101 clusters (2.41 pc) is similar to that of the MW GCs (2.98 pc) while the faint M101 clusters have a much larger median size (4.27 pc).}
\label{radhisto}
\end{figure}

Figure~\ref{radhisto} shows the histogram of cluster sizes ($r_{\text{eff}}$) for the bright and faint cluster subsamples.  The FWHM of each cluster was measured using the ISHAPE software, as described in $\S2.2$, then converted to $r_{\text{eff}}$ assuming a distance to M101 of 6.4~Mpc \citep{sha11}.  The figure shows that the median $r_{\text{eff}}$ values for the bright (2.41~pc) versus faint (4.27~pc) clusters are different, with fainter clusters tending to have larger sizes.  MW GCs have a median $r_{\text{eff}}$ of 2.98~pc, similar to that found for the bright clusters in M101.

\begin{figure}[htp]
\includegraphics[width=0.48\textwidth]{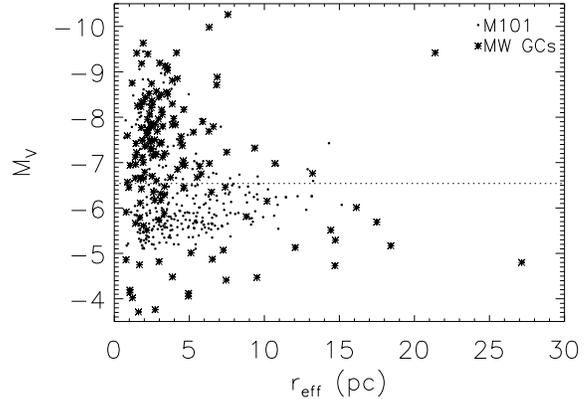}
\caption{Absolute magnitude vs. $r_{\text{eff}}$ for our M101 catalog (solid circles) and MW GCs (asterisks).  The horizontal dotted line represents the luminosity division at $M_V = -6.54$ imposed on our sample (see $\S 3.1$).  The density of the faint M101 clusters is much greater than that of the faint MW clusters.  (Although the MW clusters are a more complete sample down to fainter magnitudes.)  The brighter clusters have less spread in sizes and skew toward being more compact for both the MW and M101.}
\label{lumrad}
\end{figure}

Figure~\ref{lumrad} plots cluster luminosity versus $r_{\text{eff}}$ for the bright and faint M101 samples, and for MW GCs.  The MW GCs cover a similar parameter space as those in M101, with the bright clusters in M101 being more compact with less scatter to larger radii than the faint clusters.  

\begin{figure}[htp]
\includegraphics[width=0.48\textwidth]{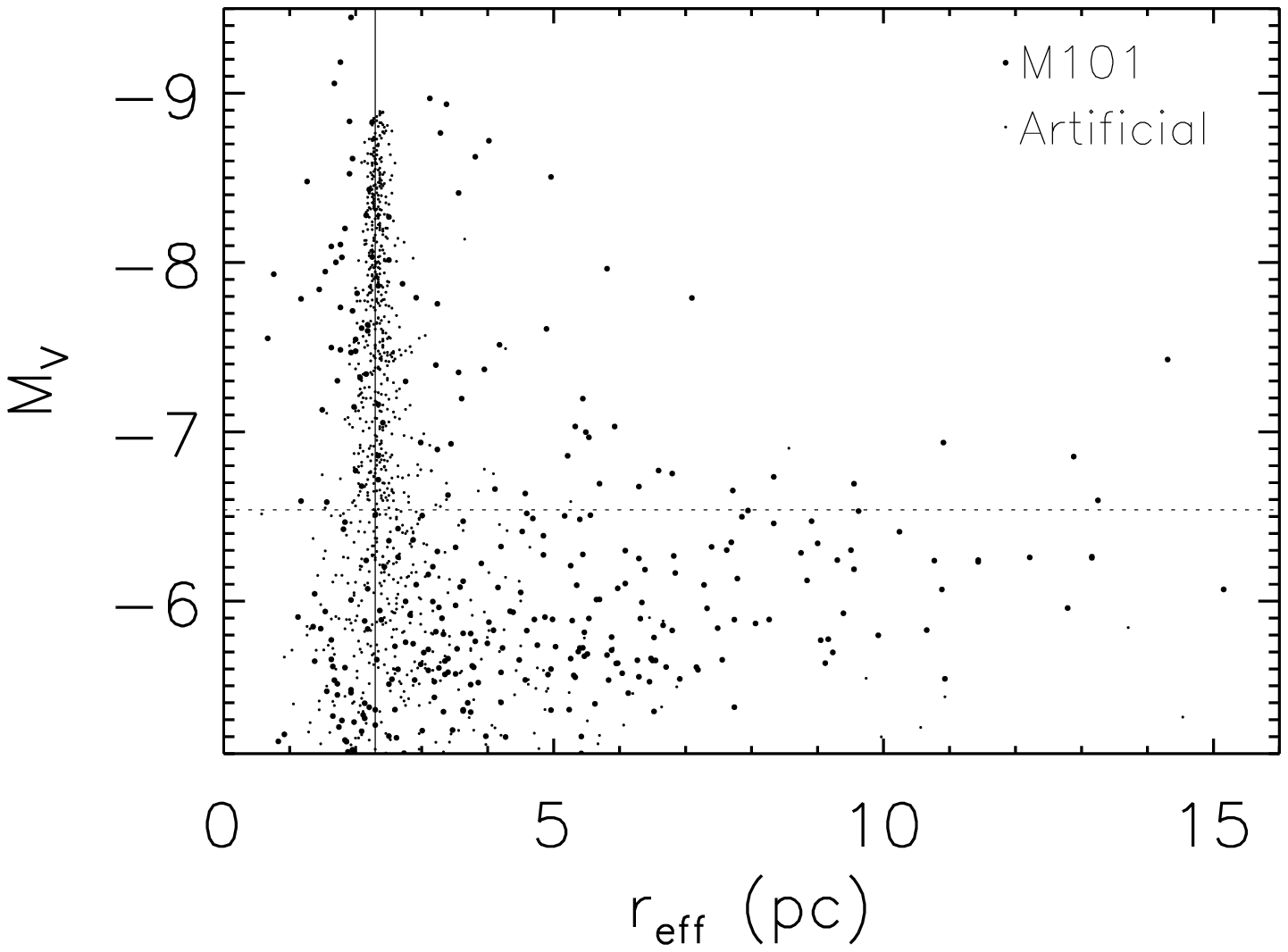}
\includegraphics[width=0.48\textwidth]{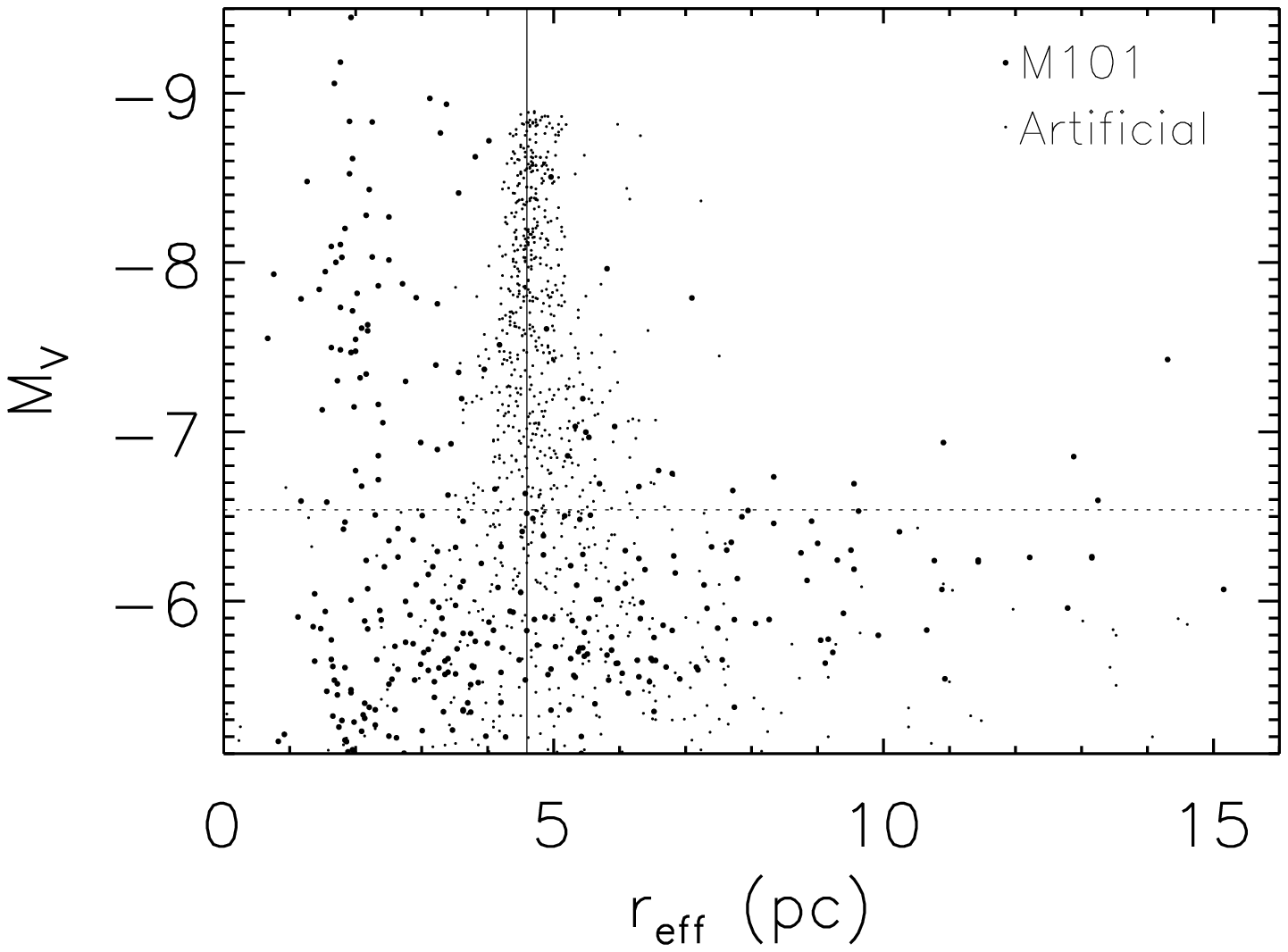}
\caption{ The same as Fig. \ref{lumrad} for the M101 clusters (large solid circles) and artificially generated clusters (small solid circles).  The vertical solid lines are the input sizes of the artificial clusters (top FWHM$=1.0$ and bottom FWHM$=2.0$).  In the top plot, the real data are reasonably well matched by artificial clusters at bright magnitudes, but at faint magnitudes, it becomes apparent that too many real clusters have large sizes that cannot be explained by the spread due to measurement errors alone.  In the bottom plot, we see that the faint clusters are better described by larger input size artificial clusters.}
\label{lumrad_art}
\end{figure}

In order to investigate whether or not there are systematic biases in the sizes measured by ISHAPE for bright versus faint clusters, we compare ISHAPE measurements of artificial clusters of varying input sizes added to the M101 images (see $\S2.3$ for details).  We use 2000 of the artificial clusters from the completeness testing with FWHM $= 1.0$ and $2.0$~pixels in addition to 2000 artificial clusters with FWHM $= 0.5$ and $4.0$~pixels.

Figure \ref{lumrad_art} plots the cluster luminosity versus measured $r_{\text{eff}}$ for the M101 clusters and the FWHM$=1.0$ and 2.0~pixel ($r_{\text{eff}}=2.3$~pc and 4.6~pc) artificial clusters.  While it is apparent from the artificial clusters that there is a spread in the measured $r_{\text{eff}}$ which increases at fainter magnitudes, it is also apparent that this spread cannot fully describe the number of real, faint clusters with large sizes.

We compare the results from artificial clusters with the observations using the Hodges-Lehmann (HL) estimator, which determines the shifts between the location parameters of two data sets, in order to quantitatively establish whether the observed difference in cluster sizes between the bright and faint subsamples is real or the result of systematic and random errors.  The results from the statistical tests and artificial cluster experiments described below suggest that there is a physical difference in the sizes of bright and faint clusters in M101.

The HL shift estimates the difference between the median values of two data sets if the data sets are symmetric about the median, or the difference between the pseudo-medians if the data sets are not symmetric about the median.  It does this by computing the median of the differences between each pair of the values in the two data sets.  It also gives confidence intervals and $p$-values for the significance of the separation as part of the Wilcoxon rank sum test.  The HL estimator indicates a shift between the median sizes of bright versus faint clusters in M101 of $1.26 \substack{+0.53 \\ -0.51}$ (errors are 95\% confidence intervals).  For the same magnitude ranges, there is a smaller shift of $0.30 \substack{+0.07 \\ -0.09}$ ($0.55 \substack{+0.18 \\ -0.21}$) for bright versus faint artificial clusters with input FWHM values of 1.0~pixels or $r_{\text{eff}}=2.3$~pc (FWHM$=2.0$, $r_{\text{eff}}=4.6$~pc).  We also tested the HL estimators for bright versus faint clusters for two faint binnings that excluded the faintest clusters and therefore highest uncertainty FWHM measurements (faint bins from $-5.6 > M_V > -6.54$ and $-6.1 > M_V > -6.54$; bright bins still $M_V \le -6.54$).  This only increased the difference between the real and artificial cluster shift estimators.  

$P$-values for the real and artificial clusters for all binnings show that the shifts are significant with values ranging from $6.2 \times 10^{-14}$ to 0.03 (strongly accepting the alternative hypothesis that the true location shift is not equal to 0).  We conclude that there is a physical difference between the sizes of bright and faint clusters in M101, with the latter tending to be larger (by $\sim0.71-0.96$~pc).

\subsection{Spatial Distribution}

\begin{figure}[htp]
\includegraphics[width=0.48\textwidth]{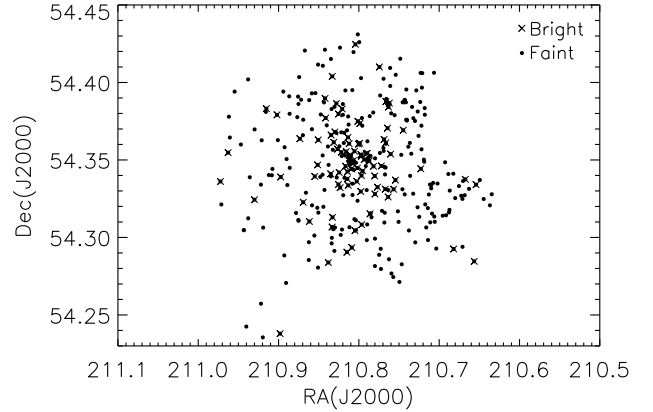}
\caption{Positions in R.A. and decl. for our cluster sample.  The bright clusters (crosses) are more centrally concentrated than the faint clusters (solid circles).}
\label{positions}
\end{figure}

\begin{figure}[htp]
\includegraphics[width=0.48\textwidth]{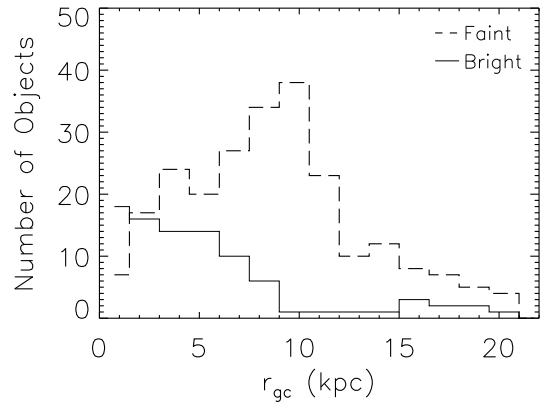}
\caption{Histogram of $r_{\text{gc}}$ for each cluster in our sample.  The bright clusters (solid line) are concentrated toward the center while the faint clusters (dashed line) have a broader distribution.}
\label{disthisto}
\end{figure}

The locations of our clusters within M101 in R.A. and decl. are shown in Figure~\ref{positions}.  Figure~\ref{disthisto} shows the $r_{\text{gc}}$ distribution for the faint versus bright clusters.  The bright clusters are largely found within 9~kpc of the center, and quite centrally concentrated, while the faint clusters are more evenly distributed.

\begin{figure}[htp]
\includegraphics[width=0.48\textwidth]{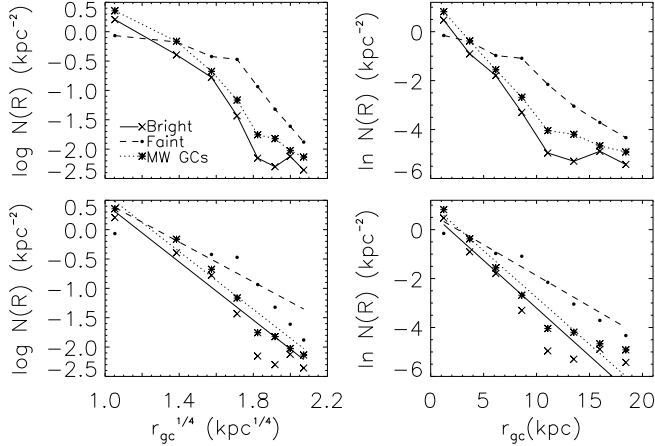}
\caption{Surface density plots showing the area normalized number of clusters at different binned radii from the center of the host galaxy in log $N$-$r_{\text{gc}}^{1/4}$ (left) and $ln N$-$r_{\text{gc}}$ space (right).  The symbols show the mid point of the radii bin they represent with either connecting lines to guide the eye (top) or best fit lines over laid (bottom).  In all plots, the bright M101 clusters (crosses with solid line) follow a centrally concentrated distribution, similar to the MW GCs (asterisks with dotted line), while the faint M101 clusters (solid circles with dashed line) are much less centrally concentrated.  Slopes of the best fits to the data are shown in Table~\ref{slopes}.}
\label{surfacedensity}
\end{figure}

%% Table of the slopes for the best fit lines to the surface density profiles for M101

%% Only uncomment the following lines (and the end{document} command at the end) if the table needs to be %% generated on its own.
%\documentclass{aastex}
 %\begin{document}

\begin{deluxetable*}{lrrrrrr}
\tablecolumns{7}
%% \rotate
\tablewidth{0pc}
\tablecaption{Coefficients for Surface Density Fits}
\tablehead{
\colhead{Data Set} & \colhead{$b_{\text{Vauc}}$} & \colhead{$c_{\text{Vauc}}$} & \colhead{$R_e$ (kpc)\tablenotemark{a}} & \colhead{$b_{\text{exp}}$} & \colhead{$c_{\text{exp}}$} & \colhead{$R_0$ (kpc)\tablenotemark{b}}}
\startdata
Bright M101 & 2.91 $\pm0.21$ & -2.47 $\pm0.15$ & 3.28 $\pm0.80$ & 0.68 $\pm0.16$ & -0.39 $\pm0.02$ & 2.57 $\pm0.13$ \\
Faint M101 & 2.17 $\pm0.19$ & -1.70 $\pm0.11$ & 14.78 $\pm3.81$ & 0.64 $\pm0.14$ & -0.25 $\pm0.02$ & 3.97 $\pm0.32$ \\
MW GCs & 3.07 $\pm0.17$ & -2.46 $\pm0.12$ & 3.35 $\pm0.66$ & 1.05 $\pm0.13$ & -0.38 $\pm0.02$ & 2.61 $\pm0.14$ \\

%Bright M101 & 4.12 $\pm0.28$ & -3.07 $\pm0.18$ & 1.38 $\pm0.22$ & 1.17 $\pm0.18$ & -0.39 $\pm0.02$ & 2.57 $\pm0.02$ \\
%Faint M101 & 2.90 $\pm0.23$ & -2.05 $\pm0.13$ & 6.91 $\pm0.16$ & 0.95 $\pm0.16$ & -0.25 $\pm0.02$ & 3.97 $\pm0.02$ \\
%MW GCs & 4.25 $\pm0.23$ & -3.04 $\pm0.15$ & 1.44 $\pm0.17$ & 1.52 $\pm0.15$ & -0.38 $\pm0.02$ & 2.61 $\pm0.02$ \\
\enddata
\tablenotetext{a}{Effective radii calculated with the formula $R_e = (-3.33/c_{Vauc})^4$ \citep{deV78}.}
\tablenotetext{b}{Scale radii calculated with the formula $R_0 = -1/c_{exp}$.}
\label{slopes}
\end{deluxetable*}

%\end{document}

Figure~\ref{surfacedensity} shows the surface density plots in $log N$-$r_{\text{gc}}^{1/4}$ and $ln N$-$r_{\text{gc}}$ space.  $N$ for the surface density plots is derived by counting the number of clusters within the annulus inner and outer radii covered by the $r_{\text{gc}}$ bin.  $N$ is then divided by the area of the annulus to determine the cluster surface density.  Here, $r_{\text{gc}}$ and $N$ do not include the innermost $10''$ ($\sim310$~pc at the distance of M101) of the galaxy since this region is too bright to detect clusters down to the same magnitude limit as the rest of the galaxy.  Beyond this region, the background level drops sharply and incompleteness does not vary strongly with $r_{\text{gc}}$.  This can be seen in Figure \ref{disthisto_art} which shows the completeness fraction as a function of $r_{\text{gc}}$.

We fit a de Vaucouleurs law of the form $log N = b_{\text{Vauc}} + c_{\text{Vauc}}r_{\text{gc}}^{1/4}$ and an exponential of the form $ln N = b_{\text{exp}} + c_{\text{exp}}r_{\text{gc}}$ to each of the three data sets which are reported in Table~\ref{slopes} (errors are Poisson errors based on the count in $N$).  The de Vaucouleurs surface brightness profile is typically used to model spheroidal components and elliptical galaxies \citep{deV48}, while exponential profiles provide a good description of disk components of galaxies \citep{pat40}.  We also convert the slopes of the de Vaucouleurs fits into effective radii ($R_e$), the radius within which half of the total cluster population lies in projection, with the formula $R_e = (-3.33/c_{Vauc})^4$ \citep{deV78}.  The slopes of the exponential fits are converted into scale radii ($R_0$) by $R_0 = -1/c_{exp}$.

Both the $R_e$ and $R_0$ are almost identical for the bright M101 clusters and MW GCs whereas the values for the faint M101 clusters differ significantly from those of the MW GCs.  Furthermore, the faint M101 clusters do not appear to follow a straight line in $log N$-$r_{\text{gc}}^{1/4}$ space (see the left panels of Fig.~\ref{surfacedensity}), and so may not be well represented by the de Vaucouleurs law.  However, they follow a straighter line path on the $ln N$-$r_{\text{gc}}$ space surface density plot, indicating an exponential law better describes the distribution of the faint clusters.  The bright clusters appear to follow a slightly straighter line in the $log N$-$r_{\text{gc}}^{1/4}$ space than $ln N$-$r_{\text{gc}}$ space, as expected from the high degree of steepness in linear space, and hence, the de Vaucouleurs law gives a better fit.

\section{Discussion}

In this section, we first examine the properties (luminosity, color, size, and spatial distributions) of the bright versus faint clusters to determine if our sample consists of two distinct populations of old star clusters.  Then we more closely examine the properties of the faint clusters and compare to those in other galaxies.

\subsection{Are There Two Populations of Old Clusters in M101?}

The shape of the LF gives the first indication that there may be two populations of red clusters in M101.  As seen in Figure~\ref{lumhisto}, the LF rises nearly continuously from the bright end down to the completeness limit, but shows a dip near $M_V\sim-7.5$ to $-6.54$.  This dip can be explained if the LF is a combination of a peaked distribution (as seen for old GCs) at the bright end and a rising, power law-like distribution (as seen for younger clusters) at the faint end.  For old GC systems in the MW and other galaxies, both luminosity and mass distributions have similar, peaked shapes, believed to result from mass-loss over $\sim12$~Gyr, mostly driven by the effects of two-body relaxation.

When we compare the MW GCLF to that of our M101 cluster sample (see Fig.~\ref{lumhisto} and $\S3.1$), we find the shapes quantitatively match well at the bright end, at least down to $M_V = -7.5$, 0.2~mag brighter than the peak in the MW GCLF \citep{ash95}.  The CMD (see Fig.~\ref{CMD}) shows that the colors of the majority of the bright clusters are similar to those of MW GCs.  The median $r_{\text{eff}}$ of the bright clusters is similar to the median size of the MW GCs (but smaller than that of the faint clusters; see \S3.3).  The bright M101 clusters are centrally concentrated toward the center of the galaxy, as are the MW GCs (see \S3.4), which is expected for a spherically distributed population.

The similarity between the luminosity, magnitude, color, size, and spatial distributions of the bright M101 clusters and the MW GCs strongly suggests that the former are most likely a typical old ($\sim12$~Gyr) population of  GCs in M101.  Furthermore, our results suggest the M101 bright clusters are largely metal-poor, consistent with the galaxy's nearly bulge-less morphology.  Figure~\ref{colorhisto} shows that the bright cluster colors skew toward the blue peak of BCGs \citep{har06}.  Corrections to the photometry that put our clusters on the same photometric system as the BCG data would push the colors further blueward as would any unaccounted for reddening corrections.  Thus, the few clusters near the red peak of the BCGs, most likely do not account for a separate metal-rich population of clusters as seen in more bulge-dominated/spheriodal galaxies.

Turning to the faint clusters, their LF can be described by a rising power law with an average slope of $\alpha=-2.6 \pm 0.3$ (see \S3.1 and Fig.~\ref{lumhisto}).  This is similar to what is found for young cluster populations in other galaxies.  \citet{lar02} found that young star cluster LFs in six spirals follow a power law of $dN(L_V)/dL_V \propto L_V^{\alpha}$ where $\alpha$ ranges from $\sim-2.0$ to $-2.6$.  \citet{whi14} examined the LFs of clusters in 20 nearby ($4-30$~Mpc) star-forming galaxies, and found an average $\alpha = -2.37$ with rms scatter of $0.18$.

In checking for similarities between the faint clusters and the MW GCs, we find the median $r_{\text{eff}}$ of the MW GCs is smaller than that of the faint clusters, and their spatial distributions are very different.  While the MW GCs and bright red M101 clusters are concentrated to the center of their host galaxies, the faint clusters in M101 are more uniformly distributed as expected if they reside in the disk.  {\bf We conclude that the faint, red clusters form a population distinct from the typical old GCs, one which is associated with the disk of M101.}

\subsection{What Are the Faint, Red Clusters in M101?}

Discovering the true nature of the faint clusters depends on determining their ages, metallicities, and masses.  It is clear from Figure~\ref{colors} that exact ages and metallicities cannot be determined accurately for each cluster from optical colors alone since the SSP tracks lie on top of each other in this color space.  Nonetheless, we can compare the colors, luminosities, and sizes of these clusters with those in other galaxies where the cluster properties are better known.  

The faint clusters may be a population similar to the LMC intermediate age clusters or the brightest old open clusters since they share a similar color-magnitude space (see Fig.~\ref{CMD}), although the faint M101 clusters do extend to brighter magnitudes than the old open clusters.  The faint fuzzies discovered by \citet{bro02} and \citet{lar00} in lenticular galaxies have $V-I \sim1.0-1.5$ which is similar to that of the faint clusters, $\sim1.1$.  Despite the similar colors and magnitudes between faint fuzzies and our faint clusters, only 48 of the 230 faint clusters with FWHM measurements have $r_{eff}>7$~pc, but the faint clusters do still have a large median size, greater than the bright clusters or MW GCs.

We can use the observed power law shape of the LF to constrain the ages of the faint clusters.  As mentioned previously, the peaked shape found for old GCs is driven by stellar mass loss due to internal relaxation.  Figure 3 in \citet{fal01}, shows how a peak develops at the low end of the mass function and then moves to higher masses as the population ages, while the high mass end continues to have a power law shape.  Since we do not observe a peak in the LF of the faint, red clusters, it must occur below our completeness limit.  When luminosities are converted to masses, the age at which the peak of the mass function is just below the completeness limit is the maximum age of the faint clusters.  

We make the following assumptions for our calculations: a stellar-mass loss rate for each cluster of $\mu_{\text{ev}}=10^{-5}$~$M_{\sun}yr^{-1}$, a typical value for MW GCs (e.g., \citet{fal01}), and a single age $\tau$ for the cluster population.  This gives a predicted peak mass, $M_\text{p}=\mu_{\text{ev}}\tau$, at different ages.  We then use the SSP model predicted, age-dependent $M/L_V$ to convert our observed cluster luminosities to masses at different assumed ages, and determine the mass equivalent of our completeness limit, $M_{\text{lim}}$.  Based on this methodology, the maximum age for our clusters is constrained by the value where $M_{\text{p}} = M_{\text{lim}}$.  

We find the maximum age to be $\sim700$~Myr ($\sim1$~Gyr) for $Z=0.008$ ($Z=0.02$).  Since the faint clusters are found in the disk, we might expect the maximum age based on the more metal-rich model to give a better estimate, but MW old open clusters span metallicities from just above solar down to [Fe/H]$\sim -1$ \citep{fri95}.  Likewise, faint fuzzies have typical [Fe/H]$\sim -0.6$ \citep{bro02}, and LMC intermediate age clusters have [Fe/H]$\sim-0.4$ to $-0.7$ \citep{pal13}.  

The faint M101 clusters do have a lower density than typical MW GCs, and the mass-loss rate due to relaxation-driven evaporation depends on the internal density of the clusters, $\mu_{\text{ev}} \propto \rho_\text{h}^{1/2}$, where $\rho_\text{h}$ is the half-light density \citep{cha07,mcl08}.  Therefore, a better estimate of the maximum age of the faint M101 clusters might be determined by scaling the mass loss rate to the median size of the faint clusters ($\mu_{\text{ev,M101}} = \mu_{\text{ev,MW}}(\rho_{\text{h,M101}}/\rho_{\text{h,MW}})^{1/2}$).   Now, we find the maximum age to be $\sim9-10$~Gyr for $Z=0.008$ and $\sim12-13$~Gyr for $Z=0.02$.  Thus it is possible that the faint clusters might be quite old.

We can also put constraints on the minimum age of the faint clusters.  \citet{bar06} suggested that the faint, red cluster population in M101 may be reddened, young disk clusters; however, they do not provide any estimated ages.  We examine $B-I$ color images which highlight the locations of dust lanes in M101 and find no preference for the faint clusters to be embedded in dust.  Likewise, we examine archival  $HST/WFPC2$ H$\alpha$ (F656N) images and find no preference for the faint clusters to be in or near \ion{H}{2} regions.  This strongly indicates that they are older than $10$~Myr, as clusters younger than this are expected to have not yet fully dispersed, leaving the gas and dust clouds from which they formed.

Also, we have $Gemini$/GMOS spectra for 25 YMC clusters which exhibit strong Balmer lines with no H$\alpha$ emission lines, indicating ages of a few hundred Myr.  We find that their median colors from our $HST$ data are $B-V\sim0.2$ and $V-I\sim0.5$ (see Fig.~\ref{colors}), significantly bluer than the faint red clusters, which have $B-V\sim0.7$ and $V-I\sim1.1$.  \citet{bar06} selected 1260 ``blue" clusters in M101 with $(B-V)_0 <0.45$ and $V<23$, similar to our spectroscopically confirmed YMCs, and they found these clusters appear to coincide with the spiral structure of the M101 disk.  The faint red clusters likely do not follow the spiral structure (despite being in the disk), because they have had sufficient time to disperse throughout the disk and away from their birth sites.  We therefore believe that the faint red clusters are older than a few hundred Myr. 

The faint red clusters are associated with the disk of M101, just as old open clusters are in the MW disk\citep{por10}.  Note that the scale radius from the exponential fit to the faint M101 clusters ($\sim4$~kpc) is similar to the scale radius of the disk of M101 according to optical photometry ($R_0=$4.6-4.8~kpc \citet{oka76}).  Interestingly, faint fuzzies have also been found to be associated with the disks of their lenticular host galaxies \citep{bro02,chi13,for14} despite their ages $\ge 7-8$~Gyr.  \citet{chi13} find that the strong association of the NGC 1023 faint fuzzies with their galaxy's disk--rather than its bulge--to be evidence that the faint fuzzies are simply very old open clusters.  They further predict that such clusters should be found in spiral galaxy disks as well, and that  the only reason they have thus far only been identified in lenticular galaxies is because of their smooth disks, which make them easier to observe.  The excellent resolution of $HST$/ACS allows us to identify such clusters in M101 for the first time.  \citet{sch07} also found six clusters in the spiral galaxy M51 that match the definition of faint fuzzies.  {\bf We conclude that the faint clusters studied here are old ($\tau \ga 700$~Myr) and part of the disk of M101, similar to but more massive than old open clusters in the MW.}

\section{Conclusions}

M101 appears to have two populations of old star clusters: a typical population of old GCs and a fainter population of intermediate-age to old disk clusters.  For the population of old GCs, we find:

\begin{enumerate}
\item Their luminosity distribution, colors, sizes, and spatial distribution are similar to those of the MW GCs.
\item Their spatial distribution shows a central concentration which is consistent with a spherically distributed halo population. 
\item Their colors are skewed toward the typical blue peak of BCGs which indicates they are most likely a metal-poor dominated population, fitting with the nearly bulge-less morphology of M101. 
\end{enumerate}

For the fainter population, we find:

\begin{enumerate}
\item Their luminosity distribution is similar to the power law shape of young cluster populations in other spiral galaxies, $dN(L_V)/dL_V \propto L_V^{\alpha}$ where typically $\alpha\approx-2$ to $-2.6$.  We find the average power law fit to our faint clusters in M101 has a slope of $\alpha=-2.6 \pm 0.3$.
\item Age constraints determined from the shape of the LF indicate that they could be quite old, up to $\sim12-13$~Gyr, although this constraint depends on a number of assumptions.
\item They are older than a few hundred Myr because their colors are much redder than a sample of M101 YMCs with spectroscopic ages of a few hundred Myr.  These clusters also do not follow the spiral arm structure that the blue M101 clusters identified by \citet{bar06} do.
\item They have a fairly extended spatial distribution, quite different from the centrally concentrated, bright GCs.  They are most likely associated with the disk.
\item They occupy the same luminosity-color space as LMC intermediate age clusters, the brightest old open MW clusters, and faint fuzzies.  Old open clusters and faint fuzzies are also located in the disks of their galaxies.  \citet{chi13} concluded that faint fuzzies are analogous to old open clusters and should be found in spiral galaxies.  We conclude the M101 faint clusters are most likely these old disk clusters.
\end{enumerate}

The evidence for a large population of old disk clusters in M101 shows that the peak of the GCLF may not be an accurate distance indicator for all galaxy types, especially not spiral and lenticular galaxies.  We would caution against using it as a stand alone measure of distance.  Other spiral galaxies need to be examined for faint, disk cluster populations.

We are in the process of using $Gemini$ GMOS spectra of bright, old GCs in M101 to measure metallicities for calibrating the color-metallcity relationship for M101.  Such a relationship can be applied to our bright cluster sample to determine more accurate ages and metallicities than can be found from the optical colors alone.  Unfortunately, we do not have spectra for any clusters in the faint population.  Preliminary results show that the bright cluster subset are quite old, with ages similar to those of Galactic GCs.  Results from this analysis will be presented in a follow-up paper.

%% Bibliography
\bibliography{ArticleFile_ArXivversion.bib}
\bibliographystyle{apj}

\end{document}